\documentclass[preprint,showpacs,aps,prb,psfig]{revtex4}
\usepackage{graphicx}

\begin{document}

\title{Spin transitions in time-dependent regular and random magnetic fields.}
\author{V.L. Pokrovsky$^{1,2}$, N.A. Sinitsyn$^1$}
\address{$^1$Department of Physics, Texas A\&M University, College Station, TX 77843-4242\\
$^2$Landau Intitute for Theoretical Physics, Chernogolovka, Moscow Distr, 142432, Russia }
\date{\today }


\begin{abstract}
We study the transition between Zeeman levels of an arbitrary spin placed
into a regular time-dependent magnetic field and a random field with the
Gaussian distribution. One component of the regular field changes its sign
at some moment of time, whereas another component
does not change substantially. The noise is assumed to be fast. In this
assumption we find analytically the ensemble average of the spin density
matrix and its fluctuations.
\end{abstract}
\maketitle

\pagenumbering{arabic}


\section{Introduction}

Nanomagnets $Fe_{8}$ and $Mn_{12}$ attracted much attention last
decade. It was shown experimentally that they realize a single
molecule quantum hysteresis \cite{wernsdorfer}. Some features of
the hysteresis have been predicted in many theoretical works
\cite{{dobrovitski},{saito},{prokofiev}}. In particular, it is
commonly accepted that plateaus on the hysteresis loop are due to
the Landau-Zener (LZ) transitions \cite{{landau},{zener}} at
avoided crossing of Zeeman-split spin levels in the crystal field
of molecular environment. This point of view is strongly supported
by the experimental observation \cite{wernlz} of theoretically
predicted \cite{garg} oscillations of the transition matrix
elements \ vs. magnetic field applied in the hard direction.
However, it was indicated earlier \cite{prokofiev} that the
hyperfine interaction with the numerous nuclei in the molecule is
comparatively large and can remarkably complicate a simple LZ
picture of transition. The reason why it did not happen at a
specific experimental setup \cite{wernlz} was recently explained
theoretically \cite{prok-sin}. It is indicated in the same work
that at a smaller rate of the field sweeping the violation of the
simple LZ formula seem to be unavoidable. The nuclear spins
provide not only a random static field which changes locally
Zeeman splitting and create a non-orthogonality. Their
fluctuations at a little higher temperature become fast and can be
considered as a random time-dependent magnetic field acting on the
spin. Another source of the noise is the interaction with phonons
\cite{chud}, which produces random, time-dependent anisotropy.

Recently several nanomagnets with cubic or almost cubic symmetry
were synthesized. One of them $\mathrm{Mo}_{6}\mathrm{Mn}_{9}$
\cite{decurtis} has spin 51/2 (this value can be comparatively
easy varied by simple substitutions). Another nanomagnet $\left[
\left( \mathrm{trifos}\right) Re\left( \mathrm{CN}\right)
_{3}\right] _{4}\left[ \mathrm{CoCl}\right] _{4}$ \cite{dunbar},
which we will abbreviate as $\mathrm{Re}_{4}\mathrm{Co}_{4},$ has
a smaller spin $S=6$. So far no effects of anisotropy were found
in magnetic measurements even at temperature about 1K. It shows
that the cubic anisotropy is weak, as it could be expected. Thus,
the cubic nanomagnets may be the best realization of a large free
spin. Being placed into a varying magnetic field $B_{z}(t)$ along
$z-$ axis and a small constant field $B_{x}$ in $x-$ direction,
such a spin performs quantum transitions between the states of the
Zeeman multiplet. The essential difference between this situation
and the standard Landau-Zener (LZ) problem
\cite{{landau},{zener}} is that at $B_{z}(t)=0$ all $2S+1$
states of the Zeeman multiplet cross simultaneously, whereas
Landau and Zener considered only two-level crossing equivalent to
$S=1/2$. An extension of the LZ theory to the case of higher spins
was proposed by Hioe \cite{hioe}. He has proved that the
transition probabilities depend only on $S$ and the dimensionless
LZ parameter
\begin{equation}
\gamma =\sqrt{g\mu _{B}B_{x}^{2}/4\hbar \dot{B}_{z}},
\label{gamma}
\end{equation}
where $g$ is the Lande factor, $\mu _{B}$ is the Bohr's magneton
and $\dot {B}_{z}$ is the time derivative (sweeping rate) of the
magnetic field taken at the diabatic level crossing, i.e. at the
moment when $B_{z}(t)$ turns into zero. The transition
probabilities display very interesting oscillations vs. the
parameter $\gamma $ and difference of projections $m$ and
$m^{\prime }$ of the orbital moment in the initial and final
states. These oscillations, which were not analyzed in the work
\cite{hioe}, originate from the quantum interference of different
Feynman paths leading from $m$ to $m^{\prime }$, which does not
exist in the genuine LZ problem \cite{PS0}. Thus, the cubic
nanomagnets are simple, but non-trivial physical objects. The
quantum tunneling in these objects is worthwhile of studying. As
it was shown for other nanomagnets, the interaction with nuclear
spins is essential for the dynamics. Though the nuclear relaxation
times are sufficiently long in microscopic scale (typically
milliseconds), the characteristic time of the LZ process $\tau
_{LZ}\sim B_{x}/ \dot {B}_{z}$ may be even longer. Then the field
of nuclear spins can be considered as a fast gaussian noise. In
the opposite case this field should be considered as a quasistatic
random field. Such a case was studied theoretically by Prokof'ev
and Stamp \cite{prokofiev}.

The purpose of our article is to study the influence of the fast
noise onto probabilities of transitions between the states of
Zeeman multiplet for a free spin $S$ in the presence of a regular
time-dependent field, the same as that in the LZ or Hioe problem.
A special case of this problem for $S=1/2$, when the regular field
$B_{x}$ is zero and transitions are completely determined by
non-diagonal elements of the random field, was solved earlier by
Kayanuma \cite{kayanuma}, who has found average values of the
transition probabilities. In our previous work \cite{PS1} we have
solved the same problem for regular and random fields acting
together. In this work we extend our results for higher spins,
find the average non-diagonal matrix elements of the density
matrix and calculate the fluctuations of all these values, which
occur to be strong. These calculations became possible due to high
symmetry of the problem. A proper group-theoretical treatment
allows to deal with standard objects, which we call Bloch tensors.
They are a generalization of the well-known Bloch vector for spin
1/2 problem.

The plan of the article is as follows. In section 2 we remind the
Hioe solution and analyze in some details the oscillations of the
transition probabilities. In section 3 we formulate equations for
the density matrix and reduce them to equations for the Bloch
tensors, which we define in the same section. In section 4 we
consider the fast noise acting on a two-level system. Though this
problem was considered in our previous article \cite{PS1}, it is
important to give a simple analysis of the time scales and to
calculate average transition matrix and fluctuations, which will
serve as a basis for the spin-$S$ problem. Section 5 is dedicated
to the solution of the same problem for general spin. In section
6, returning to 2-level system (spin 1/2), we discuss the limit of
a strong regular field $B_{x}$, so that the regular transition
proceeds adiabatically. The noise remains fast in the scale
$\tau_{LZ}$, but its relaxation frequency may be much larger than
$B_x$. The last 7-th section contains our conclusions.

\section{Spin $S$ Landau-Zener problem}

The Hamiltonian of a free spin $\mathbf{S}$ with the maximal projection $S$ in an
external time-dependent magnetic field $\mathbf{B}\left( t\right) $ reads:

\begin{equation}
H_{S}(t)=-\mathbf{Sb}\left( t\right)  \label{HS}
\end{equation}

\noindent where $\mathbf{b}\left( t\right) =g\mu _{B}\mathbf{B}\left(
t\right) $. The key observation \cite{{hioe},{PS0}} is that this
Hamiltonian is an operator of time-dependent infinitesimal rotation.
Therefore the corresponding evolution matrix $U_{S}(t,t_{0})=T\exp \left(
-i\int\limits_{t_{0}}^{t}H_{S}(t^{\prime })dt^{\prime }\right) $ is an
operator of rotation belonging to the group $SO(3)$ acting in its
irreducible representation labeled by an integer or half-integer $S$. Since
the composition law is the same for any irreducible representation, the
resulting evolution operator represents the same rotation for any spin. The
group theory allows to construct this matrix for an arbitrary spin if it is
known for spin 1/2 (see \cite{LL}, ch. VIII). The multi-spinor technique is
most appropriate for this purpose. The spin $S$ state can be represented as
a direct symmetrized product of $2S$ spin 1/2 states:
\begin{equation}
\left\vert S,m\right\rangle =\sqrt{\frac{(S+m)!(S-m)!}{(2S)!}}\left(
\left\vert ++...+--...-\right\rangle +\left\vert ++...-+...-\right\rangle
+...\right)   \label{multispinor}
\end{equation}%
where each ket contains $S+m$ spins up (+) and $S-m$ spins down ($-$) and
all permutations of up and down are performed. Let the $SU(2)$ matrix
rotating spin 1/2 states be:

\begin{equation}
u=\left(
\begin{array}{cc}
a & b \\
-b^{\ast } & a^{\ast }%
\end{array}%
\right)  \label{SU2}
\end{equation}%
with \ the constraint $|a|^{2}+|b|^{2}=1$ imposed. The transformation for
the state (\ref{multispinor}) of the spin $S$ can be obtained as the direct
product of transformations (\ref{SU2}):

\bigskip

\bigskip
\begin{equation}
\begin{array}{c}
\left\vert S,m\right\rangle \rightarrow \sqrt{\frac{(S+m)!(S-m)!}{(2S)!}}%
a^{S+m}(-b^{\ast })^{S-m}\left\vert S,S\right\rangle + \\
\left( \frac{\sqrt{2S}(2S-1)!}{(S-m-1)!(S+m)!}a^{S+m}(-b^{\ast
})^{S-m-1}a^{\ast }+\frac{\sqrt{2S}(2S-1)!}{(S+m-1)!(S-m)!}%
a^{S+m-1}(-b^{\ast })^{S-m}b\right) \left\vert S,S-1\right\rangle ...%
\end{array}
\label{smss}
\end{equation}
A general matrix element of the rotation operator $\left\langle m\right\vert
U_{S}\left\vert m^{\prime }\right\rangle $ for the spin $S$ is expressed in
terms of $a,b,a^{\ast },b^{\ast }$ in the following way \cite{{LL},{vilenkin}}:
\begin{equation}
\left\langle m\right\vert U_{S}\left\vert m^{\prime }\right\rangle =\left[
\frac{(S+m^{\prime })!(S-m^{\prime })!}{(S+m)!(S-m)!}\right]
^{1/2}a^{m^{\prime }+m}b^{m^{\prime }-m}P_{S-m^{\prime }}^{m^{\prime
}-m,m^{\prime }+m}\left( 2|a|^{2}-1\right)   \label{general}
\end{equation}%
where $P_{n}^{a,b}(x)$ are the Jacobi polynomials \cite{erdelyi}. The matrix
elements possess the following symmetry properties: $\left\langle
-m\right\vert U_{S}\left\vert -m^{\prime }\right\rangle
=(-1)^{|m|+|m^{\prime }|}\left\langle m\right\vert U_{S}\left\vert m^{\prime
}\right\rangle ^{\ast }$, \ $\left\vert \left\langle m\right\vert
U_{S}\left\vert m^{\prime }\right\rangle \right\vert =\left\vert
\left\langle m^{\prime }\right\vert U_{S}\left\vert m\right\rangle
\right\vert =\left\vert \left\langle -m\right\vert U_{S}\left\vert
-m^{\prime }\right\rangle \right\vert $. Equation (\ref{general}) displays
oscillations of the matrix element when the argument $2|a|^{2}-1$ varies
from $-1$ to +1. These oscillations are associated with the oscillatory
behavior of the Jacobi Polynomial. For the number of nodes $N(m,m^{\prime
},S)$ of the matrix element $\left\langle m\right\vert U_{S}\left\vert
m^{\prime }\right\rangle $ a simple equation is valid: $N(m,m^{\prime
},S)=S-\max (|m|,|m^{\prime }|)$. The central matrix element with $%
m=m^{\prime }=0$ for integer $S$ and $|m|=|m^{\prime }|=1/2$ for
half-integer $S$ has maximal number of nodes equal to $S$ and $S-1/2$,
respectively.

Let us specify the problem considering only a narrow vicinity of the
diabatic levels crossing point, which we accept for $t=0$. It is possible if
the interval of time $\tau _{LZ}$, during which the transitions presumably
proceed is much less than the characteristic time of the field variation $%
t_{0}=\left| \dot {B}_{z}/ \ddot {B}_{z}\right| $.
The characteristic time of transition can be identified as $\tau
_{LZ}=\left| B_{x}/\dot{B}_{z}\right| $. Thus, the requirement
of short transition time can be rewritten as $B_{x}\ll \left( \dot
{B}_{z}\right) ^{2}/\left|  \ddot{B}_{z}\right| $. If this
requirement is satisfied, one can approximate with high accuracy the
magnetic field by a linear function of time $B_{z}(t)=\dot{B}_{z}%
t$.

Landau and Zener \cite{{landau},{zener}} have solved such a problem for
spin 1/2. In particular Zener have determined matrix elements $a,b$ in terms
of the Weber function $D_{-i\gamma ^{2}}(e^{i\pi /4}\sqrt{\dot {
\omega }t)}$ for an arbitrary moment of time. For simplicity we will focus
on the values of these parameters for transition from $t=-\infty $ to $%
t=\infty $. According to the Landau-Zener solution
\begin{equation}
a=\exp \left( -\pi \gamma ^{2}\right) ;\quad b=-\frac{\sqrt{2\pi }\exp
\left( -\frac{\pi \gamma ^{2}}{2}+\frac{i\pi }{4}\right) }{\gamma \Gamma
(-i\gamma ^{2})}  \label{LZab}
\end{equation}
When $\gamma $ varies from 0 to $\infty $, the modulus $|a|$ changes from $1$
to 0 and the argument of the Jacobi polynomial in equation (\ref{general})
varies from 1 to $-1$. Using the expression (\ref{general}), one finds the
solution for an arbitrary spin $S$ in terms of coefficients $a$ and $b$, or
equivalently, in terms of the Landau-Zener parameter $\gamma $. Thus, the
transition amplitudes oscillate as function of $\gamma $. The physical
reason of these oscillations is the interference between different Feynman
paths leading from $m$ to $m^{\prime }$. For illustration we show here
corresponding matrices for spins 1, 3/2 and 2:

\begin{equation}
U_{1}=\left(
\begin{array}{ccc}
a^{2} & -\sqrt{2}ab & -b^{2} \\
\sqrt{2}ab^{*} & 2|a|^{2}-1 & \sqrt{2}a^{*}b \\
-b^{*2} & -\sqrt{2}a^{*}b^{*} & a^{*2}%
\end{array}
\right)  \label{U1}
\end{equation}
\begin{equation}
U_{3/2}=\left(
\begin{array}{cccc}
a^{3} & \sqrt{3}a^{2}b & \sqrt{3}ab^{2} & b^{3} \\
-\sqrt{3}a^{2}b^{*} & \left( 3|a|^{2}-2\right) a & \left( 3|a|^{2}-1\right) b
& \sqrt{3}a^{*}b^{2} \\
\sqrt{3}ab^{*2} & -\left( 3|a|^{2}-1\right) b^{*} & \left( 3|a|^{2}-2\right)
a^{*} & \sqrt{3}a^{*2}b \\
-b^{*3} & \sqrt{3}a^{*}b^{*2} & -\sqrt{3}a^{*2}b^{*} & a^{*3}%
\end{array}
\right)  \label{U3/2}
\end{equation}
We present only a quarter of the transition matrix for $S=2$. The rest can
be found by using above described symmetry properties.

\begin{equation}
U_{2}=\left(
\begin{array}{ccccc}
a^{4} & \sqrt{4}a^{3}b & \sqrt{6}a^{2}b^{2} & \sqrt{4}ab^{3} & b^{4} \\
& \left( 4|a|^{2}-3\right) a^{2} & \sqrt{6}\left( 2|a|^{2}-1\right) ab &
\left( 4|a|^{2}-1\right) b^{2} &  \\
&  & 6|a|^{4}-6|a|^{2}+1 &  &  \\
&  &  &  &  \\
&  &  &  &
\end{array}
\right)  \label{U2}
\end{equation}

\section{Density matrix and Bloch tensors.}

When the random magnetic field acts onto the spin, the system must be described by
the density matrix $\widehat{\rho }$. By definition it is a $(2S+1)\times
(2S+1)$ Hermitian matrix with the trace equal to 1. It satisfies the
standard equation of motion:

\begin{equation}
i\frac{d\widehat{\rho }}{dt}=\left[ H,\widehat{\rho }\right]  \label{rho}
\end{equation}
Any Hermitian matrix with the trace equal to 1 can be represented as a sum:

\begin{equation}
\begin{array}{l}
\widehat{\rho } =\frac{1}{2S+1}I+\mathbf{g\cdot S+}\frac{1}{2}g_{ik}\left(
S_{i}S_{k}+S_{k}S_{i}-\frac{2}{3}\delta _{ik}S(S+1)\right) +   \\
...\frac{1}{\left( 2S\right) !}g_{i_{1}i_{2}...i_{2S}}\left(
S_{i_{1}}S_{i_{2}}...S_{i_{2S}}+\mathrm{all~permutations-all~traces}\right)
\end{array}
\label{tensors}
\end{equation}
If the Hamiltonian is the generator of the rotation (\ref{HS}), each term in
equation (\ref{tensors}) corresponds to an irreducible representation and
evolves independently. We will call symmetric tensors $g_{ik}$, $g_{ikl}$ ...%
$g_{i_{1}i_{2}...i_{2S}}$ the Bloch tensors in analogy with the Bloch vector
$\mathbf{g}$ well known from Bloch theory of the nuclear spin motion. Any
trace of such a tensor must be equal to zero. The Hamiltonian (\ref{HS})
generates following equations of motion for the Bloch tensors:

\begin{equation}
\dot{\mathbf{g}}=-\mathbf{b\times g;\quad }\dot{g}%
_{ik}=-\varepsilon _{ilm}b_{l}g_{mk}-\varepsilon _{klm}b_{l}g_{im};...
\label{t-equations}
\end{equation}
All these equations are independent and have obvious integrals of motion:

\begin{equation}
\mathbf{g}^{2}=const;~g_{ik}^{2}=const;~g_{ikl}^{2}=const;...
\label{invariants}
\end{equation}
Thus, the density matrix of a spin $S$ in an external time-dependent
magnetic field has $2S$ conserving values. It is convenient to represent the
Bloch tensors by their complex component with definite projection to the $z-$%
axis. We will denote such components of a tensor of the rank $s$ as $g_{s,m}$%
. The corresponding tensor operators composed from the symmetrized products
of $2s$ components of the spin $S$ operators are denoted $T_{s,m}^{S}$. They
can be constructed from the senior operator of this representation $%
T_{s,s}^{S}=2^{-s/2}S_{+}^{s}$ with $S_{\pm }=S_{x}\pm iS_{y}$ by recurrent
commutations with the operator $S_{-}$:
\begin{equation}
T_{s,m}^{S}=-\frac{1}{\sqrt{(s+m+1)(s-m)}}\left[ S_{-},T_{s,m+1}^{S}\right]
\label{recursion}
\end{equation}
The operators $T_{s,m}^{S}$ are polynomials of the standard spin operators $%
S_{\pm }$ and $S_{z}$. They are operator analogs of spherical harmonics.
Several lowest such operators are presented in Appendix. We show below
relations between the Cartesian components of the tensor $g_{i_{1}...i_{s}}$
and its components $g_{s,m}$ for several values of $s$:

\begin{eqnarray}
g_{1,\pm 1} &=&\frac{1}{\sqrt{2}}\left( g_{x}\pm ig_{y}\right)
;~g_{1,0}=g_{z}   \\
g_{2,\pm 2} &=&\frac{1}{\sqrt{6}}\left( g_{xx}-g_{yy}\pm 2ig_{xy}\right)
;~g_{2,\pm 1}=\frac{1}{\sqrt{2}}\left( g_{xz}\pm ig_{yz}\right)
;g_{2,0}=g_{zz}  \label{g(sm)} \\
g_{3,\pm 3} &=&\frac{1}{\sqrt{20}}\left( g_{xxx}\pm 3ig_{xxy}-3g_{xyy}\mp
ig_{yyy}\right) ;...
\end{eqnarray}
The general rule for writing the $s,\pm m-$component via its Cartesian
counterparts is the same as for the product $\frac{m!}{\sqrt{2m!}}(x\pm
iy)^{m}z^{s-m}$. The Hamiltonian (\ref{HS}) in terms of components with
definite projections reads (note that $b_{\pm }=b_{x}\pm ib_{y};\,S_{\pm
}=S_{x}\pm iS_{y})$:
\begin{equation}
H=-b_{z}S_{z}-\frac{1}{2}(b_{+}S_{-}+b_{-}S_{+})  \label{HS-proj}
\end{equation}
Equations (\ref{t-equations}) in terms of the components with definite $z-$%
projections read:

\begin{equation}
\dot {g}_{s,m}=-imb_{z}g_{s,m}+\frac{i}{2}\sqrt{(s+m)(s-m+1)}%
b_{+}g_{s,m-1}+\frac{i}{2}\sqrt{(s-m)(s+m+1)}b_{-}g_{s,m+1}  \label{g(sm)-eq}
\end{equation}%
and the conservation laws are:

\begin{equation}
\sum\limits_{m=-s}^{s}\left| g_{s,m}\right| ^{2}=const  \label{invariants-m}
\end{equation}

\section{Fast noise in two-level system\label{fast2}}

In this section we consider only spin 1/2 or, equivalently a two-level
system. We assume that the magnetic field can be separated into regular and
random parts:

\begin{equation}
\mathbf{b(}t\mathbf{)=b}_{r}(t)+\mathbf{\eta (}t\mathbf{)}
\label{separation}
\end{equation}
where $\mathbf{b}_{r}(t)=\hat{z}\dot{b}_{z}t+%
\hat{x}b_{x}$ and $\mathbf{\eta (}t\mathbf{)}$ is the
Gaussian noise determined by its correlators:

\begin{equation}
\left\langle \eta _{i}(t)\eta _{k}(t^{\prime })\right\rangle
=f_{ik}(t-t^{\prime })  \label{correlator}
\end{equation}
We assume that the correlators (\ref{correlator}) decay after a
characteristic time difference $\tau _{n}$ and that this correlation time is
much less than the characteristic time of the LZ process $\tau _{LZ}$.
However, the noise must be slow enough to avoid the direct transitions
between the levels when the interlevel distance approaches its saturation or
characteristic value $\omega $ far from the crossing point. Thus, the noise
correlation time $\tau _{n}$ must satisfy a following inequalities:

\begin{equation}
\omega ^{-1}\ll \tau _{n}\ll \tau _{LZ}  \label{inequality}
\end{equation}
The spectral width of noise is $1/\tau _{n}$. The noise produces transitions
during the interval of time $t_{acc}=1/(\dot {b}_{z}\tau _{n})$,
after which the current LZ frequency becomes larger than the noise spectral
width. We will call this interval the accumulation time and assume that it
is much larger than other characteristic time intervals $\tau _{n}$ and $%
\tau _{LZ}$.

We first solve an auxiliary problem in which $b_{x}=0$ and transitions are
mediated by noise only. Such a problem for a special shape of correlators ($%
f_{xx}=J^{2}\exp \left( -\frac{t-t^{\prime }}{\tau _{n}}\right) $; the
remaining components of the correlation tensor are zero) was solved earlier
by Kayanuma \cite{kayanuma} and studied numerically by Nishino \textit{et al.%
} \cite{nishino}. In our work \cite{PS1} we have generalized and simplified
the Kayanuma solution. Here we reproduce our solution \cite{PS1} in a
modified form convenient for extension to higher spins. We also obtain new
results calculating the fluctuation of the density matrix, or equivalently
the Bloch vector. Equations for the component of the Bloch vector in this
case are:
\begin{equation}
\dot {g}_{z}=(i/\sqrt{2})\left( \eta _{+}g_{-}-\eta
_{-}g_{+}\right) ;\,\,\dot{g}_{\pm }=\mp i\left( \dot
{b}_{z}t+\eta _{z}\right) g_{\pm }+(i/\sqrt{2})\eta _{\pm }g_{z}
\label{eq-half}
\end{equation}%
Solving equation for $g_{\pm }$, we find:

\begin{eqnarray}
g_{\pm }(t) &=&g_{\pm }(-\infty )\exp \left( \mp \frac{i\dot {b}%
_{z}t^{2}}{2}\mp i\int\limits_{-\infty }^{t}\eta _{z}(t^{\prime })dt^{\prime
}\right) +   \\
&&(i/\sqrt{2})\int\limits_{-\infty }^{t}\exp \left[ \mp \frac{i
\dot {b}_{z}(t^{2}-t^{\prime 2})}{2}\mp i\int\limits_{t^{\prime }}^{t}\eta
_{z}(t^{\prime \prime })dt^{\prime \prime }\right] \eta _{\pm }(t^{\prime
})g_{z}(t^{\prime })dt^{\prime }  \label{gpm}
\end{eqnarray}%
Let us first consider the case of complete initial decoherence: $g_{\pm
}(-\infty )=0$. Then, plugging equation (\ref{gpm}) into the first equation (%
\ref{eq-half}), we find a separate equation for $g_{z}$:
\begin{equation}
\dot {g}_{z}=-(1/2)\int\limits_{-\infty }^{t}\exp \left[ -\frac{i%
\dot{b}_{z}(t^{2}-t^{\prime 2})}{2}-i\int\limits_{t^{\prime
}}^{t}\eta _{z}(t^{\prime \prime })dt^{\prime \prime }\right] \eta
_{+}(t)\eta _{-}(t^{\prime })g_{z}(t^{\prime })dt^{\prime }+c.c.
\label{gz-before}
\end{equation}%
Let us average equation (\ref{gz-before}) over the ensemble of the random
noise. For such averaging it is important that the noise correlation time $%
\tau _{n}$ is much shorter than the time $t_{acc}$ necessary for a
substantial variation of $\left\langle g_{z}\right\rangle $. This fact
allows to represent the average $\left\langle \eta _{+}(t)\eta
_{-}(t^{\prime })g_{z}(t^{\prime })\right\rangle $ approximately as a
product: $\left\langle \eta _{+}(t)\eta _{-}(t^{\prime })g_{z}(t^{\prime
})\right\rangle =\left\langle \eta _{+}(t)\eta _{-}(t^{\prime
})\right\rangle \left\langle g_{z}(t^{\prime })\right\rangle $. More
accurately one should incorporate the fluctuations of $g_{z}$. In the
leading approximation they are determined by the same equation (\ref%
{gz-before}) as follows:
\begin{equation}
\delta g_{z}=-(1/2)\int\limits_{-\infty }^{t}dt_{1}\int\limits_{-\infty
}^{t_{1}}\exp \left[ -\frac{i\dot{b}_{z}(t_{1}^{2}-t_{2}^{2})}{2}%
\right] \left( \eta _{+}(t_{1})\eta _{-}(t_{2})-\left\langle \eta
_{+}(t_{1})\eta _{-}(t_{2})\right\rangle \right) \left\langle
g_{z}(t_{2})\right\rangle dt_{2}+c.c  \label{gz-fluct}
\end{equation}%
We ignore $\eta _{z}$ (this approximation will be justified by the next
step). Let substitute this additional term into equation (\ref{gz-before})
and perform averaging over the gaussian random field $\mathbf{\eta }$.
According to the Wick's rule, it is reduced to all possible pairings. In our
case the only possible pairing is $\left\langle \eta _{+}(t)\eta
_{-}(t_{2})\right\rangle \left\langle \eta _{-}(t^{\prime })\eta
_{+}(t_{1})\right\rangle $. Such a pairing limits the integration by the
interval $t-\tau _{n}<t_{2}<t_{1}<t^{\prime }<t$. Thus the contribution of
the fluctuational term differs by an additional factor $\sim \tau
_{n}/t_{acc}<<1$ from the principal contribution from $\left\langle
g_{z}\right\rangle $. These arguments represent a shortened version of the
original arguments by Kayanuma \cite{kayanuma} and are akin to the
Abrikosov-Gor'kov theory of static disordered alloys \cite{AG}.

Using the fact that the decay of the correlator $\left\langle \eta
_{+}(t)\eta _{-}(t^{\prime })\right\rangle $ limits effectively the
integration over time by an interval $t-\tau _{n}<t^{\prime }<t$, we can
prove that the contribution of the noise component $\eta _{z}$ in the
exponent in equation (\ref{gz-before}) can be neglected. To estimate this
contribution we assume that $\eta _{z}$ is statistically independent from
other components. Then the averaging over $\eta _{z}$ results in the
Debye-Waller factor $\exp \left[ -\frac{1}{2}\left\langle \left(
\int\limits_{t^{\prime }}^{t}\eta _{z}(t^{\prime \prime })dt^{\prime \prime
}\right) ^{2}\right\rangle \right] $. The argument of this exponent can be
estimated as $\left\langle \eta _{z}^{2}\right\rangle \tau _{n}^{2}$ . It is
small provided the level of noise $\left\langle \eta _{z}^{2}\right\rangle $
is much smaller than $\tau _{n}^{-2}$. The noise correlators are even
functions of the time difference. Therefore, expanding linearly the time
argument of the exponent in the same equation $\frac{\dot{b}%
_{z}(t^{2}-t^{\prime 2})}{2}\approx \dot {b}_{z}t(t-t^{\prime })$%
, one can transform the integral-differential equation (\ref{gz-before})
into an ordinary differential equation for $\left\langle g_{z}\right\rangle $%
:
\begin{equation}
\dot{\left\langle g_{z}\right\rangle }=-\hat{F}\left( \dot{b}_{z}t\right) \left\langle g_{z}\right\rangle  \label{diff}
\end{equation}
where $\hat{F}(\Omega )$ is the Fourier-transform of the function $F(\tau
)=f_{xx}(\tau )+f_{yy}(\tau )$:
\begin{equation}
\hat{F}(\Omega )=\int\limits_{-\infty }^{\infty }F(\tau )\cos \Omega \tau
d\tau  \label{Fourier-cos}
\end{equation}

\noindent Equation (\ref{diff}) has a simple solution:

\begin{equation}
\left\langle g_{z}(t)\right\rangle =\left\langle g_{z}(-\infty
)\right\rangle \exp \left[ -\int\limits_{-\infty }^{t}\hat{F}\left(
\dot{b}_{z}t^{\prime }\right) dt^{\prime }\right]  \label{gzav(t)}
\end{equation}
At $t\rightarrow +\infty $ the asymptotic value of $\left\langle
g_{z}\right\rangle $ is:

\begin{equation}
\left\langle g_{z}(+\infty )\right\rangle =\left\langle g_{z}(-\infty
)\right\rangle \exp \left( -\theta \right) ;\quad \theta =\frac{\pi F(0)}{%
\dot{b}_{z}}  \label{gzas}
\end{equation}%
Note that what matters for the LZ transition is the average quadratic
fluctuation of non-diagonal noise at any moment $F(0)=\left\langle \eta
_{x}^{2}+\eta _{y}^{2}\right\rangle $ in contrast to a standard
characteristic of the white noise which would be $\hat{F}(0)$. Indeed,
commonly white noise correlator is introduced as $\left\langle \eta (t)\eta
(t^{\prime })\right\rangle =\gamma \delta (t-t^{\prime })$. The only
characteristic of the noise in this approach is $\gamma
=\int\nolimits_{-\infty }^{\infty }\left\langle \eta (t)\eta (t^{\prime
})\right\rangle dt^{\prime }$. An interesting feature of the asymptotic
formula (\ref{gzas}) is its independence on the noise correlation time $\tau
_{n}$. However, it should be kept in mind that this asymptotic is valid only
at time $t\gg t_{acc}=\left( \dot{b}_{z}\tau _{n}\right) ^{-1}.$

Returning to the solution (\ref{gpm}) for $g_{\pm }$, we see that the term $%
g_{\pm }(-\infty )\exp \left( \mp \frac{i\dot{b}_{z}t^{2}}{2}\mp
i\int\limits_{-\infty }^{t}\eta _{z}(t^{\prime })dt^{\prime }\right) ,$
omitted at substitution in the first equation (\ref{eq-half}) and the
ensemble averaging, vanishes if the $z-$component of the noise is
statistically independent from others. Now let us perform a similar
procedure solving first equation for $g_{z}$ and then substituting the
solution into equations for $g_{\pm }$. In the same approximation equation
for averages of these components of the Bloch vector reads:

\begin{equation}
\left\langle \dot{g}_{\pm }\right\rangle =-\frac{1}{2}\hat{F}%
\left( \dot{b}_{z}t\right) \left\langle g_{\pm }\right\rangle
\label{gpmav}
\end{equation}%
The first term in equation (\ref{gpm}) after the averaging turns into zero
at any finite $t$. Indeed the Debye-Waller factor which appears in this case
is $\exp \left[ -\frac{1}{2}\left\langle \left( \int\limits_{t}^{\infty
}\eta _{z}(t^{\prime \prime })dt^{\prime \prime }\right) ^{2}\right\rangle %
\right] =0$. The asymptotics of $\left\langle g_{\pm }\right\rangle $ at $%
t\rightarrow +\infty $ is:
\begin{equation}
\left\langle g_{\pm }(+\infty )\right\rangle =\exp \left( -\frac{\theta }{2}%
\right) \left\langle g_{\pm }(-\infty )\right\rangle ;\quad \theta =\frac{%
\pi F(0)}{\dot {b}_{z}}  \label{gpmas}
\end{equation}%
Note that the symbols $g_{\pm }(\pm \infty )$ denote the coefficients at $%
\exp \left( \mp \frac{i\dot{b}_{z}t^{2}}{2}\right) $. Thus, the
noise asymptotically tends to reduce the average components of the Bloch
vector, i.e. to establish equipopulation of the levels and to destroy the
coherence. However, during the time interval of the order of $t_{acc\text{ }%
} $the average Bloch vector can oscillate.

For the considered problem it is possible to find exactly the fluctuations
of the Bloch vector. Indeed, they are given by a standard formula:

\begin{equation}
\left\langle \left( \delta \mathbf{g}\right) ^{2}\right\rangle =\left\langle
\mathbf{g}^{2}\right\rangle -\left\langle \mathbf{g}\right\rangle ^{2}
\label{someequatin}
\end{equation}
Due to the symmetry of the problem, $\mathbf{g}^{2}$ is a conserving value.
Therefore, its average coincides with itself and is determined by
initial conditions. The average value of the vector $\left\langle \mathbf{g}%
\right\rangle $ was calculated above. Thus, we find an asymptotic value of
the fluctuations:

\begin{equation}
\left. \left\langle \left( \delta \mathbf{g}\right) ^{2}\right\rangle
\right\vert _{t=+\infty }=g_{z}^{2}(-\infty )\left( 1-\exp \left( -2\theta
\right) \right) +\left[ g_{x}^{2}(-\infty )+g_{y}^{2}(-\infty )\right]
\left( 1-\exp \left( -\theta \right) \right)   \label{fluct}
\end{equation}%
The values of average square fluctuation can be also written for any moment
of time. The fluctuations are strong, i.e. their magnitudes are of the same
order as the average values of the Bloch vector components unless $\theta $
is very small. An important property of the noise is that in fluctuations it
mixes diagonal and non-diagonal elements of the density matrix, i.e all
three components of the Bloch vector.

Let us proceed to the solution of a more general problem retaining non-zero $%
x-$component of the regular field $b_{x}$. Such a solution becomes possible
due to separation of times \cite{PS1}: we can neglect the effect of $b_{x}$
beyond the LZ time interval $\tau _{LZ}$ near the crossing point. On the
other hand, we can neglect the effect of the noise inside this and even much
larger interval of time, since its characteristic time is $t_{acc}\gg \tau
_{LZ}$. Thus, the problem is separated into 3 parts: in two intervals $%
\left( -\infty ,-t_{0}\right) $ and $\left( t_{0},+\infty \right) $ we can
use the solution of our auxiliary problem, i.e. to take in account only
transitions caused by the noise; in the interval $\left( -t_{0},t_{0}\right)
$ the Landau-Zener solution is valid. We need only to match them properly.
The time interval $t_{0}$ must satisfy a strong inequality: $\tau _{LZ}\ll
t_{0}\ll t_{acc}$. As we have shown before, the average components $g_{z}$
and $g_{\pm }$ evolve separately under the action of the noise. Therefore, at the moment of time $-t_{0}$ these components are:

\begin{equation}
\left\langle g_{z}(-t_{0})\right\rangle =\exp (-\theta /2)g_{z}(-\infty
);\quad \left\langle g_{\pm }(-t_{0})\right\rangle =\exp (-\theta /4)g_{\pm
}(-\infty )  \label{-t0}
\end{equation}%
The action of the LZ transition matrix (\ref{SU2}) with the matrix elements (%
\ref{LZab}) can be transferred onto the vector $\mathbf{g}$ by using spin-1
matrix (\ref{U1}). Thus, the average components of the Bloch vector at $%
t=t_{0}$ are as follows,

\begin{eqnarray}
\left\langle g_{z}(t_{0})\right\rangle &=&\sqrt{2}ab^{*}\left\langle
g_{+}(-t_{0})\right\rangle +(2|a|^{2}-1)\left\langle
g_{z}(-t_{0})\right\rangle +\sqrt{2}a^{*}b\left\langle
g_{-}(-t_{0})\right\rangle  \label{+t0z} \\
\left\langle g_{+}(t_{0})\right\rangle &=&a^{2}\left\langle
g_{+}(-t_{0})\right\rangle -\sqrt{2}ab\left\langle
g_{z}(-t_{0})\right\rangle -b^{2}\left\langle g_{-}(-t_{0})\right\rangle
\label{+t0+} \\
\left\langle g_{-}(t_{0})\right\rangle &=&-b^{*2}\left\langle
g_{+}(-t_{0})\right\rangle -\sqrt{2}a^{*}b^{*}\left\langle
g_{z}(-t_{0})\right\rangle +a^{*2}\left\langle g_{-}(-t_{0})\right\rangle
\label{+t0-}
\end{eqnarray}
where $a$ and $b$ are given by equations (\ref{LZab}). The transition from $%
+t_{0}$ to $+\infty $ is provided by the same diagonal transition matrix (%
\ref{-t0}), which was already used for the transition from $-\infty $ to $%
-t_{0}$:

\begin{equation}
\left\langle g_{z}(+\infty )\right\rangle =\exp (-\theta
/2)g_{z}(+t_{0});~\left\langle g_{\pm }(+\infty )\right\rangle =\exp
(-\theta /4)g_{\pm }(+t_{0})  \label{+infty}
\end{equation}
Collecting together equations (\ref{-t0}-\ref{+infty}), we find the final
result:

\begin{eqnarray}
\left\langle g_{z}(+\infty )\right\rangle &=&\sqrt{2}\exp \left( -3\theta
/4\right) \left( ab^{*}g_{+}(-\infty )+a^{*}bg_{-}(-\infty )\right) +\exp
\left( -\theta \right) (2|a|^{2}-1)g_{z}(-\infty )  \label{+inftyz} \\
\left\langle g_{+}(+\infty )\right\rangle &=&\exp \left( -\theta /2\right)
\left( a^{2}g_{+}(-\infty )-b^{2}g_{-}(-\infty )\right) -\exp \left(
-3\theta /4\right) \sqrt{2}abg_{z}(-\infty )  \label{+infty+} \\
\left\langle g_{-}(+\infty )\right\rangle &=&\exp \left( -\theta /2\right)
\left( -b^{*2}g_{+}(-\infty )+a^{*2}g_{-}(-\infty )\right) -\exp \left(
-3\theta /4\right) \sqrt{2}a^{*}b^{*}g_{z}(-\infty )  \label{+infty-}
\end{eqnarray}
Let us analyze first the case of complete decoherence at $t=-\infty $, i.e. $%
g_{\pm }(-\infty )=0$. Then equations (\ref{+inftyz}-\ref{+infty-}) look
much simpler:

\begin{eqnarray}
\left\langle g_{z}(+\infty )\right\rangle  &=&\exp \left( -\theta \right)
(2|a|^{2}-1)g_{z}(-\infty )  \label{decohz} \\
\left\langle g_{+}(+\infty )\right\rangle  &=&-\exp \left( -3\theta
/4\right) \sqrt{2}abg_{z}(-\infty )  \label{decoh+}
\end{eqnarray}%
Equation (\ref{decohz}) shows that, in the absence of initial coherence the
population difference can only decrease after the transition. The noise only
strengthens this tendency. However, the initial coherence if exists can
increase the difference of population to the value larger than initial.
Equation (\ref{decoh+}) demonstrates an inverse process: the coherence can
appear after the transition even if it was absent in the initial state. It
is straightforward to derive the transition probability from equation (\ref%
{decohz}):

\begin{equation}
P_{1\rightarrow 2}=\frac{1}{2}\left[ 1-\exp (-\theta )(2|a|^{2}-1)\right] =%
\frac{1}{2}\left[ 1+\exp (-\theta )-2\exp (-\theta -2\pi \gamma ^{2})\right]
\label{1-2prob}
\end{equation}%
This formula was obtained in our previous work \cite{PS1}. At zero noise
intensity ($\theta =0$) this result turns into the Landau-Zener transition
probability. At very big noise ($\theta =\infty $) the probability is equal
to 1/2. The probability is the same for transitions $1\rightarrow 2$ and $%
2\rightarrow 1$. This symmetry does not look strange keeping in mind that we
considered the classical noise, which produces the induced transitions. It
vanishes as soon as the quantum nature of the noise and spontaneous transitions
are taken into account.

For completeness we present a formula for quadratic fluctuations at $%
t=\infty $ in the case of complete initial decoherence:

\begin{equation}
\left\langle \mathbf{g}^{2}(+\infty )\right\rangle -\left\langle \mathbf{g}%
(+\infty )\right\rangle ^{2}=\left| g_{z}(-\infty )\right| ^{2}\left[
1-e^{-2\theta }-4\left( e^{-2\pi \gamma ^{2}}-e^{-4\pi \gamma ^{2}}\right)
\left( e^{-3\theta /2}-e^{-2\theta }\right) \right]  \label{fluctuations}
\end{equation}
The fluctuations vanish at $\theta =0$ and reach their maximum value equal
to $\left| g_{z}(-\infty )\right| ^{2}$ at $\theta =\infty $.

\section{Fast noise at a multilevel crossing.}

We consider only the case of a Zeeman multiplet placed into a varying
magnetic field. It is described by equations (\ref{g(sm)-eq}). The averaging
of them over the fast noise is performed by two steps as it was done in the
previous section. First we neglect the transitions produced by the regular
part of the magnetic field and take in account only the transitions produced
by the random field. This approach is correct outside the time interval $%
\tau _{LZ}$ near the avoided crossing point $t=0$. Assuming for
simplification complete initial decoherence, we find for the average
diagonal matrix elements following equations:

\begin{equation}
\left\langle g_{s,0}^{S}(+\infty )\right\rangle =\exp \left[ -\frac{s\left(
s+1\right) }{2}\theta \right] \left\langle g_{s,0}^{S}(-\infty )\right\rangle
\label{gs}
\end{equation}
Note that these relations do not contain $S$ explicitly, the time evolution
depends on $s$ only. In the course of derivation of equation (\ref{gs}) we
ignored not only the fluctuations of $g_{s,0}^{S}$, but also all higher
components except of $g_{s,\pm 1}^{S}$. Their contributions to the main
component $g_{s,0}^{S}$ have the same order of magnitude $\sim \tau
_{n}/t_{acc}$ as the contribution of fluctuations.

The generalization of equation (\ref{gs}) to the higher projections $m$,
i.e. to the coherence factors requires some care. The truncated system of
equations (\ref{g(sm)-eq}) for zero initial values of all
components of the $s$-tensor except of $g_{s,m}^{S}$, reads

\begin{eqnarray}
\dot{g}_{s,m}^{S} &=&-im\dot{b}_{z}tg_{s,m}^{S}+i\left( \lambda _{s,m}\eta
_{+}g_{s,m-1}^{S}+\lambda _{s,-m}\eta _{-}g_{s,m+1}^{S}\right) ;
\label{trunk0} \\
\dot{g}_{s,m-1}^{S} &=&-i\left( m-1\right) \dot{b}_{z}tg_{s,m-1}^{S}+i%
\lambda _{s,m}\eta _{-}g_{s,m}^{S}  \label{trunk-} \\
\dot{g}_{s,m+1}^{S} &=&-i\left( m+1\right) \dot{b}_{z}tg_{s,m+1}^{S}+i%
\lambda _{s,-m}\eta _{+}g_{s,m}^{S}  \label{trunk+}
\end{eqnarray}
 where $\lambda _{s,m}=\sqrt{(s-m)(s+m+1)}$.
To deal with slow-varying average values the fast oscillating exponent
should be eliminated. In order to do that we introduce slow variables $%
\tilde{g}_{s,m}^{S}=g_{s,m}^{S}\exp \left( \frac{i\dot{b}_{z}t^{2}}{2}%
m\right) $. Further we consider only slow amplitudes $\tilde{g}_{s,m}$ and omit the sign tilde.
After elimination of the values $g_{s,m\pm 1}^{S}$ and
averaging, we find a following equation for $g_{s,m}^{S}$:

\begin{equation}
\left\langle \dot{g}_{s,m}^{S}\right\rangle =-\frac{1}{2}\left[ \left(
s(s+1)-m^{2}\right) \hat{F}(\dot{b}_{z}t)+m\hat{G}(\dot{b}_{z}t)\right]
\left\langle g_{s,m}^{S}\right\rangle
\end{equation}%
where $\hat{F}(\Omega )$ is defined by equation (\ref{Fourier-cos}) and $%
\hat{G}(\Omega )$ is defined as a sine Fourier-transform:
\begin{equation}
\hat{G}(\Omega )=\int\limits_{-\infty }^{\infty }\left\langle \eta _{x}(\tau
)\eta _{y}(0)-\eta _{y}(\tau )\eta _{x}(0)\right\rangle \sin \Omega \tau d\tau
\label{Fourier-sine}
\end{equation}%
Thus, the time dependence of the average $\left\langle
g_{s,m}^{S}\right\rangle $ is defined as:
\begin{equation}
\left\langle g_{s,m}^{S}(t)\right\rangle =\exp \left\{ -\frac{1}{2}%
\int\limits_{-\infty }^{t}\left[ \left( s(s+1)-m^{2}\right) \hat{F}(\dot{b}%
_{z}t^{\prime })+m\hat{G}(\dot{b}_{z}t^{\prime })\right] dt^{\prime
}\right\} \left\langle g_{s,m}^{S}(-\infty )\right\rangle   \label{coh(t)}
\end{equation}%
Its asymptotic value at $t\rightarrow +\infty $ does not contain the sine
Fourier-transform:
\begin{equation}
\left\langle g_{s,m}^{S}(+\infty )\right\rangle =\exp \left[ -\frac{1}{2}%
\left( s(s+1)-m^{2}\right) \theta \right] \left\langle g_{s,m}^{S}(-\infty
)\right\rangle   \label{coh(infty)}
\end{equation}%
At $s=1,m=1$ this result coincides with (\ref{gpmav}).

Now we proceed to our main problem including both regular and random
transverse magnetic fields. We will employ the separation of their action in
time proven in the previous section. To avoid lengthy formulas our
consideration will be restricted to the case of complete initial
decoherence. Then the only non-zero components of the Bloch tensors are $%
g_{s,0}^{S}$. Their evolution is described by three independent factors, two
of them originating from the noise and the central factor being the
generalized Landau-Zener-Hioe matrix element:
\begin{equation}
\left\langle g_{s,0}^{S}(+\infty )\right\rangle =\exp \left[ -\frac{s(s+1)}{2%
}\theta \right] P_{s}^{0,0}(2e^{-2\pi \gamma ^{2}}-1)g_{s,0}^{S}(-\infty )
\label{B-tens-fin}
\end{equation}%
where $P_{s}^{0,0}(x)$ is the Jacobi polynomial. The average values of the
Bloch tensors components with $m\neq 0$ vanish as a result of averaging over
the random phases in the initial state. To find the transition probabilities
$P_{j\rightarrow j^{\prime }}$ it is necessary to put all the diagonal
elements of the density matrix except of $\rho _{jj}$ equal to zero in the
initial state:

\begin{equation}
\frac{1}{2S+1}+\sum\limits_{s=1}^{2S}g_{s,0}^{S}(-\infty )\left(
T_{s,0}^{S}\right) _{k,k}=\delta _{jk},  \label{g-initial}
\end{equation}%
and find from these equations the initial values $g_{s,0}^{S}(-\infty )$.
Then the transition probabilities are:
\begin{equation}
P_{j\rightarrow j^{\prime }}=\frac{\delta _{jj^{^{\prime }}}}{2S+1}%
+\sum\limits_{s=1}^{2S}g_{s,0}^{S}(+\infty )\left( T_{s,0}^{S}\right)
_{j,j^{\prime }}\exp \left( -\frac{s(s+1)}{2}\theta \right)   \label{g-final}
\end{equation}%
where $g_{s,0}^{S}(+\infty )$ are defined by equation (\ref{B-tens-fin}). In
the following tables we demonstrate the results of calculations according to
this algorithm for the values of $S$ 1, 3/2 and 2 (the results for $S=1/2$
are given in the previous section by equation.(\ref{1-2prob})).

\begin{equation}
\begin{array}{llll}
_{j^{\prime }}\backslash ^{j} &\,\,\,\,\,\,\,\,\,\,\,\,\, +1 &\,\,\,\,\,\,\, 0 &\,\,\,\,\,\, -1 \\
+1 & \frac{1}{3}+\frac{1}{2}E_{1}+\frac{1}{6}E_{2} &  & \\
 0 & \frac{1}{3}-\frac{1}{3}E_{2} & \frac{1}{3}+\frac{2}{3}E_{2} &  \\
-1 & \frac{1}{3}-\frac{1}{2}E_{1}+\frac{1}{6}E_{2}  &  &
\end{array}
\label{3l}
\end{equation}

\begin{equation}
\begin{array}{lllll}
_{j^{\prime }}\backslash ^{j} &\,\,\,\,\,\, \,\,\,\,\,\, \, \, \, \, \, \, \,\, 3/2 &\,\,\,\,\,\, \, \, \, \,\,\,\,\,\,\,\, 1/2 &\,\,\,\,\,\,\,\,\, -1/2 &\,\,\, -3/2 \\
3/2 & \frac{1}{4}+\frac{9E_{1}}{20}+\frac{E_{2}}{4}+\frac{E_{3}}{20} &  &  &  \\
1/2 &
\frac{1}{4}+\frac{3E_{1}}{20}-\frac{E_{2}}{4}-\frac{3E_{3}}{20}
 &
\frac{1}{4}+\frac{E_{1}}{20}+\frac{E_{2}}{20}+\frac{9E_{3}}{20} &
 &  \\
-1/2 & \frac{1}{4}-
\frac{3E_{1}}{20}-\frac{E_{2}}{4}+\frac{3E_{3}}{20} & \frac{1}{4}-
\frac{E_{1}}{20}+\frac{E_{2}}{20}-\frac{9E_{3}}{20} &  &  \\
-3/2 & \frac{1}{4}-\frac{%
9E_{1}}{20}+\frac{E_{2}}{4}-\frac{E_{3}}{20} & & &
\end{array}
\label{4lev}
\end{equation}%
\begin{equation}
\begin{array}{llllll}
_{j^{\prime }}\backslash ^{j} &\,\,\,\,\,\,\,\,\,\,\, \, \, \, \, \, \, \, \, \, \, \, \,\,\, 2 &\,\,\,\,\,\,\,\,\,\, \, \, \, \, \, \, \, \,\,\,\,\, 1 &\,\,\,\, \,\, \,\,\,\,\,\,\,\, 0 &\, -1 &\, -2 \\
2 & \frac{1}{5}+\frac{2E_{1}}{5}+\frac{2E_{2}}{7}+\frac{E_{3}}{10}+\frac{%
E_{4}}{70} &  &  &  & \\
1 & \frac{1}{5}+\frac{E_{1}}{5}-\frac{E_{2}}{7}-\frac{E_{3}}{5}-%
\frac{2E_{4}}{35} & \frac{1}{5}+\frac{E_{1}}{10}+\frac{E_{2}}{14}+\frac{2E_{3}}{5}+\frac{%
8E_{4}}{35} &  &  &  \\
0 & \frac{1}{5}-\frac{2E_{2}}{7}+\frac{3E_{4}}{35}
& \frac{1}{5}+\frac{E_{2}}{7}-\frac{12E_{4}}{35}
& \frac{1}{5}+\frac{2E_{2}}{7}+\frac{18E_{4}}{35} &  &  \\
-1 & \frac{1%
}{5}-\frac{E_{1}}{5}-\frac{E_{2}}{7}+\frac{E_{3}}{5}-\frac{2E_{4}}{35} & \frac{1}{5}-%
\frac{E_{1}}{10}+\frac{E_{2}}{14}-\frac{2E_{3}}{5}+\frac{8E_{4}}{35} &  &  &  \\
-2 & \frac{1}{5}+\frac{2E_{1}}{5}+\frac{2E_{2}}{7}+\frac{E_{3}}{10}+\frac{E_{4}}{%
70}  &  &  &  &
\end{array}
\label{5lev}
\end{equation}%
Here we denoted $E_{s}=\exp \left( -\frac{s(s+1)}{2}\theta \right) $, i.e. $%
E_{1}=e^{-\theta };~E_{2}=e^{-3\theta };~E_{3}=e^{-6\theta
};~E_{4}=e^{-10\theta }$. Unfilled sites in the table can be easily restored
using the time reversal symmetry: $P_{j\rightarrow j^{\prime }}=P_{j^{\prime
}\rightarrow j}=P_{-j\rightarrow -j^{\prime }}$.

 In Fig1 we compare our formulas with the numerical solution of the
Sch\"odinger equation for the spin $S=1$, which convincingly confirms our
analytical results.


\begin{figure}
\includegraphics{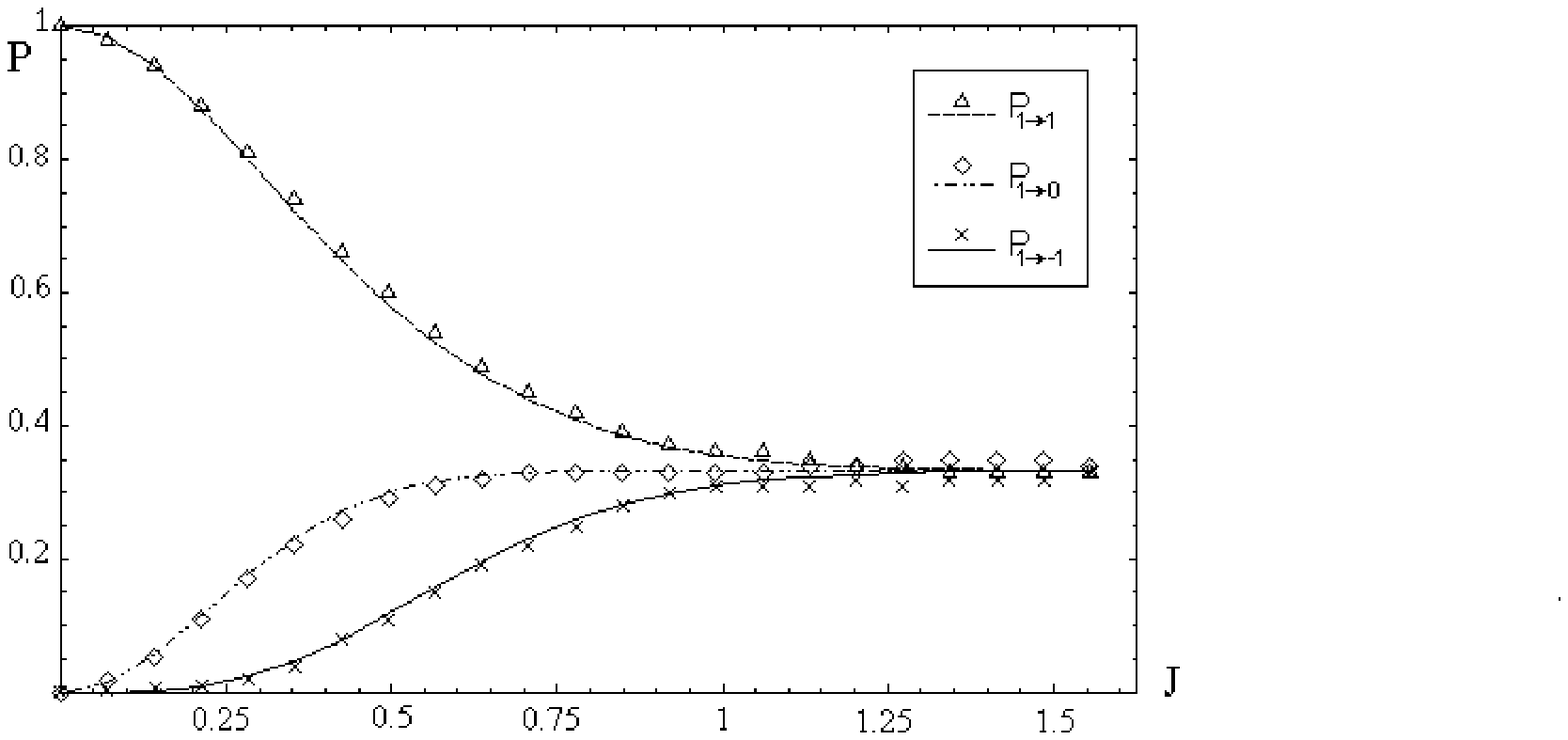}
\vskip-2in
\caption{The final probability to find a spin-1 system in the state
with $S_{z}=-1,$ $0$ or $1$ when the initial state is $S_{z}=1$ as function
of the noise amplitude. The Hamiltonian is $H=tS_{z}+\eta _{x}S_{x}$ where $%
\left\langle \eta _{x}(t_{1})\eta_{x} (t_{2})\right\rangle =J^{2}e^{-\lambda
|t_{1}-t_{2}|}$. Discrete points correspond to results of numerical
simulations with averaging over $200$ different noise realizations and $%
\lambda =125$. Lines correspond to analitical predictions of (\ref{3l}). }
\label{gfig}
\end{figure}


The quadratic fluctuations of the values $g_{s,m}^{S}$ for fixed $S$ and $s$
are calculated as it was done in the previous section:

\begin{equation}
\left\langle \delta \sum\limits_{m=-s}^{s}\left| g_{s,m}^{S}\right|
^{2}\right\rangle =\left. \sum\limits_{m=-s}^{s}\left| g_{s,m}^{S}\right|
^{2}\right| _{t=-\infty }-\left\langle g_{s,m}^{S}\right\rangle ^{2}
\label{fluct-Ss}
\end{equation}
In the case of complete initial decoherence employing equation (\ref%
{B-tens-fin}), we find:

\begin{equation}
\left\langle \delta \sum\limits_{m=-s}^{s}\left\vert g_{s,m}^{S}\right\vert
^{2}\right\rangle _{t=+\infty }=\left[ g_{s,0}^{S}(-\infty )\right] ^{2}%
\left[ 1-E_{s}^{2}\left( P_{s}^{0,0}\left( 2e^{-2\pi \gamma ^{2}}-1\right)
\right) ^{2}\right]  \label{fluct-decoh}
\end{equation}

\bigskip

\section{Noise at adiabatic level crossing.}

In this section we consider a situation of adiabatically changing levels,
i.e. $\dot{b}_{z}\ll b_{x}^{2}$, but still the noise is supposed to be
sufficiently fast $\tau _{n}\ll (\dot{b}_{z})^{-1/2}$.  The adiabatic Landau-Zener
transitions in coupled two level systems have been proposed as a candidate
for implementation of quantum gates in quantum computing \cite{bell}. It is important to understand the influence of noise
on such a q-bit manipulations in order to control the error propagation in quantum gate circuit.
We do not specify the relationship between $b_{x}$ and $\tau _{n}$.  This study is motivated by  experiments
with nanomagnets \cite{{wernlz},{V15}} in which inequality $\sqrt{\dot{b}_{z}}\tau _{n}\ll 1$ can be
easily realized since the sweeping rate of the applied magnetic field can be
made arbitrarily small. However, the noise may be not fast enough to compete
with the tunnelling amplitude $b_{x}$.  The nuclear bath
correlation time is in the range of $\tau _{n}\sim 1ms$, whereas the
measured values of the tunnelling amplitude for known nanomagnets range
between $10^{-10}-10^{-3}K$, or equivalently $10-10^{8}Hz$. In a part of
this interval $b_{x}\tau _{n}$ has a rather large value.

We consider in this section only 2-level systems. A natural approach to this
problem is the transfer to the adiabatic set of states, i.e. to the
eigenstates of the instantaneous regular part of the Hamiltonian (\ref%
{HS-proj}). Let denote this time-dependent eigenvectors as $a(t)=\left(
\begin{array}{l}
a_{1}(t) \\
a_{2}(t)%
\end{array}%
\right) $ and $b(t)=\left(
\begin{array}{l}
-a_{2}(t) \\
a_{1}(t)%
\end{array}%
\right) $, where $a_{1}(t)=\sqrt{\frac{\varepsilon (t)+b_{x}}{2\varepsilon }}%
;~a_{2}\left( t\right) =\sqrt{\frac{\varepsilon (t)-b_{x}}{2\varepsilon }}$
and $\varepsilon (t)=\sqrt{b_{z}^{2}+b_{x}^{2}}$, ($b_{z}=\dot{b}_{z}t$).
The unitary matrix of transformation to the adiabatic set reads:

\begin{equation}
U(t)=\left(
\begin{array}{ll}
a_{1}(t) & a_{2}(t) \\
-a_{2}(t) & a_{1}(t)%
\end{array}
\right) =a_{1}(t)I-ia_{2}(t)\sigma _{y}  \label{transform}
\end{equation}
where $\sigma _{\alpha }~(\alpha =x,y,z)$ are Pauli matrices. In the new
basis the total Hamiltonian acquires the following form:

\begin{equation}
H=\varepsilon (t)\sigma _{z}+U\mathbf{\eta \sigma }U^{-1}=\varepsilon
(t)\sigma _{z}+\mathbf{\eta }^{\prime }\mathbf{\sigma }
\label{Ham-transform}
\end{equation}
The transformation of the random field components is:

\begin{equation}
\eta _{x}^{\prime }=(a_{1}^{2}-a_{2}^{2})\eta _{x}+2a_{1}a_{2}\eta _{z}=%
\frac{b_{z}}{\varepsilon }\eta _{x}+\frac{b_{x}}{\varepsilon }\eta
_{z};~\eta _{y}^{\prime }=\eta _{y};~\eta _{z}^{\prime }=-\frac{b_{x}}{%
\varepsilon }\eta _{x}+\frac{b_{z}}{\varepsilon }\eta _{z}
\label{field-transform}
\end{equation}
In this form the Hamiltonian (\ref{Ham-transform}) essentially coincides
with the Hamiltonian of the auxiliary problem (zero regular transverse
field) for the two-level system (see Section \ref{fast2}). The essential
difference is first, that the effective regular external field is not linear
in time; instead it is equal to $\varepsilon (t)$; second, the correlators
of effective noise $\mathbf{\eta }^{\prime }$ now depend not only on the
time difference, but also on time itself due to the time-dependent
transformation (\ref{field-transform}):

\begin{eqnarray}
\left\langle \eta _{x}^{\prime }(t)\eta _{x}^{\prime }(t^{\prime
})\right\rangle &=&\frac{1}{\varepsilon (t)\varepsilon (t^{\prime })}\left[
\dot{b}_{z}^{2}tt^{\prime }\left\langle \eta _{x}(t)\eta _{x}(t^{\prime
})\right\rangle +b_{x}^{2}\left\langle \eta _{z}(t)\eta _{z}(t^{\prime
})\right\rangle +\right.   \\
&&\left. \dot{b}_{z}b_{x}t\left\langle \eta _{x}(t)\eta _{z}(t^{\prime
})\right\rangle +\dot{b}_{z}b_{x}t^{\prime }\left\langle \eta _{z}(t)\eta
_{x}(t^{\prime })\right\rangle \right]  \label{corr-rot}
\end{eqnarray}
Still the noise correlation time is small in comparison to the
characteristic time of variaton for the adiabatic energy $\varepsilon (t)$.
Employing the same approximation as in Section \ref{fast2}, we arrive at a
similar equation of motion for average in the case of complete initial
decoherence:

\begin{equation}
\left\langle \dot{g}_{z}^{\prime }(t)\right\rangle =-F^{\prime
}(t)\left\langle g_{z}^{\prime }(t)\right\rangle  \label{deq-rot}
\end{equation}
where
\begin{eqnarray}
F^{\prime }(t) &=&\hat{f}_{yy}\left( \varepsilon (t)\right) +\frac{1}{%
\varepsilon ^{2}(t)}\left[ \dot{b}_{z}^{2}t^{2}\hat{f}_{xx}\left(
\varepsilon (t)\right) +b_{x}^{2}\hat{f}_{zz}\left( \varepsilon (t)\right)
\right.  \\
&&+\left. \dot{b}_{z}b_{x}t\left( \hat{f}_{xz}\left( \varepsilon (t)\right) +%
\hat{f}_{zx}\left( \varepsilon (t)\right) \right) \right]
\label{Fourier-rot}
\end{eqnarray}
In the last equation the hats symbolize Fourier-transforms of corresponding
correlators. As before, we can find the average value $\left\langle
g_{z}(t)\right\rangle $ at arbitrary moment of time. Asymptotically at $%
t\rightarrow +\infty $ we find:

\begin{equation}
\left\langle g_{z}^{\prime }(+\infty )\right\rangle =\exp \left[
-\int\limits_{-\infty }^{\infty }\frac{\dot{b}_{z}^{2}t^{2}\hat{f}%
_{xx}\left( \varepsilon (t)\right) +\varepsilon ^{2}(t)\hat{f}_{yy}\left(
\varepsilon (t)\right) +b_{x}^{2}\hat{f}_{zz}\left( \varepsilon (t)\right) }{%
\varepsilon ^{2}(t)}dt\right] \left\langle g_{z}^{\prime }(-\infty
)\right\rangle  \label{asympt-adiab}
\end{equation}
The characteristic time after which the correlators in equation (\ref%
{asympt-adiab}) become very small and decay rapidly is determined by
approximate equation $\varepsilon (t)\tau _{n}\approx 1$. If $\tau _{n}\ll
b_{x}^{-1}$, then this characteristic time coincides with the accumulation
time $t_{acc}=\left( \dot{b}_{z}\tau _{n}\right) ^{-1}$ defined in Section %
\ref{fast2}, terms proportional to $b_{x\text{ }}$are negligibly small and
we return to the result (\ref{+inftyz}) with $|a|=1$, or equivalently to (%
\ref{gzas}) of the Section \ref{fast2}. In the opposite case $\tau _{n}\gg
b_{x}^{-1}$ the value $\varepsilon (t)$ exceeds $\tau _{n}^{-1}$ at any
moment of time $t$. Therefore, all correlators are small and the value of
exponent in (\ref{asympt-adiab}) is close to 1. It means that practically no
transition proceeds due to the noise between adiabatic states. Thus,
equation (\ref{asympt-adiab}) carries most interesting information when $%
\tau _{n}\sim b_{x}^{-1}$. An interesting feature of the transition
probability is that the $z$-component of noise can produce transition
between adiabatic states. This happens because the latter rotate with time.
Note that $z$-component of noise is irrelevant if $\tau _{n}\ll b_{x}^{-1}$.

\section{Conclusions}

Motivated by synthesis and magnetic measurements of cubic nanomagnets, we
developed a theory which allows to find the transition probabilities between
the states of the Zeeman multiplet in the presence of the regular
time-dependent and random magnetic field (noise). The solution of this
problem occurs to be possible since the evolution matrix for the quantum
problem is a rotation matrix acting in a spin $S$ representation. The
density matrix can be expanded into a linear superposition of irreducible
tensor operators. The coefficients at this operators related to one of the
irreducible representations (Bloch tensors) evolve independently on others.
Thus, the initial problem in the space of dimensionality $\left( 2S+1\right)
\times \left( 2S+1\right) $ is reduced to $2S$ separate problems in the
linear spaces of dimensionality from $1$ to $2S$.

The second key observation is that, for the fast noise, the transitions due to
the noise and those due to the regular part of magnetic field are separated
in time. This fact allows to solve the problems for regular field and for
the noise in the absence of the regular non-diagonal field separately and
then match them. An interesting conclusion of our theory is that, in
contrast to usual statistical calculations with the white noise, in which
only the Fourier-component of the noise correlation function with zero
frequency matters, the transition probabilities in the Landau-Zener problem
depend only on the average square of the random field amplitude. We were
able to find asymptotically exact analytical results for the probabilities.
From them we concluded, that, in the absence of initial coherence, the
average values of the diagonal components of the Bloch tensors (with zero
projection onto the direction of sweeping field) decrease monotonously with
time due to the noise. It means that the population differences in average
can only decrease after the transitions. However, they can grow if there is
a coherence in the initial state and non-diagonal components of the Bloch
tensors are not zero. Due to high symmetry the considered system has $2S$
additional integrals of motion: traces of the square of each Bloch tensor.
Thus, the increase of the population differences proceeds at the expense of
the non-diagonal components, i.e. coherence amplitudes and vice versa..

The same conservation laws enabled us to find exactly the fluctuations of
the Bloch's tensors, in particular the fluctuations of the transition
probabilities in the genuine Landau-Zener problem. They are of the same
order of magnitude that the average values.

The noise in our theory is considered as the classical random field. To
incorporate the quantum properties of noise is an interesting and
challenging problem. The second unsolved problem is to study the correlation
of the Bloch tensors at different moments of time.

\section{acknowledgement}
This work was supported by NSF under the grants DMR 0072115, DMR 0103455 and DMR 0321572,
by DOE under the grant DE-FG03-96ER45598 and by Telecommunication and
Informatics Task Force at Texas A\&M University. V.P. acknowledges the prize
from the Humboldt Foundation and Prof. Thomas Nattermann and University of
Cologne, Germany for the hospitality extended to him at final stage of this
work.


\section{Appendix}

Here we present several lowest operator spherical harmonics $T_{s,m}^{S}$.
In contrast to scalar spherical harmonics they depend on 3 parameters, two
of them are integers ($s$ and $m$), whereas $S$ accepts integer and
half-integer values. The simplest nontrivial harmonics are vectors with $s=1$%
. They are:

\begin{equation}
T_{1,0}^{S}=S_{z};~T_{1,\pm 1}^{S}=\frac{1}{\sqrt{2}}S_{\pm }
\label{vectorial}
\end{equation}
Next we demonstrate second order tensorial harmonics:

\begin{equation}
T_{2,\pm 2}^{S}=\frac{1}{2}S_{\pm }^{2};~T_{2,\pm 1}^{S}=\frac{1}{2}\left(
S_{\pm }S_{z}+S_{z}S_{\pm }\right) ;~T_{2,0}^{S}=\sqrt{\frac{3}{2}}\left[
S_{z}^{2}-\frac{1}{3}S\left( S+1\right) \right]  
\label{2-tensor}
\end{equation}
The third rank harmonics read:

\begin{eqnarray}
T_{3,\pm 3}^{S} &=&\frac{1}{2^{3/2}}S_{\pm }^{3};~T_{3,\pm 2}^{S}=\frac{1}{%
\sqrt{6}}\left( S_{\pm }^{2}S_{z}+S_{\pm }S_{z}S_{\pm }+S_{z}S_{\pm
}^{2}\right)   \\
T_{3,\pm 1}^{S} &=&\sqrt{\frac{5}{12}}\left[
S_{z}^{2}S_{+}+S_{z}S_{+}S_{z}+S_{+}S_{z}^{2}-\frac{3S(S+1)-1}{5}S_{+}\right]
\label{3-tensor} \\
T_{3,0}^{S} &=&\sqrt{\frac{5}{2}}\left[ S_{z}^{3}-\frac{3S(S+1)-1}{5}S_{z}%
\right]
\end{eqnarray}
For the fourth rank harmonics we find:

\begin{eqnarray}
T_{4,\pm 4}^{S} &=&\frac{1}{4}S_{\pm }^{4};~~   \\
T_{4,\pm 3}^{S} &=&\frac{1}{2^{5/2}}\left( S_{\pm }^{3}S_{z}+S_{\pm
}^{2}S_{z}S_{\pm }+S_{\pm }S_{z}S_{\pm }^{2}+S_{z}S_{\pm }^{3}\right)
\\
T_{4,\pm 2}^{S} &=&\frac{\sqrt{7}}{12}\left[ S_{z}^{2}S_{\pm
}^{2}+S_{z}S_{\pm }S_{z}S_{\pm }+S_{z}S_{\pm }^{2}S_{z}+S_{\pm
}S_{z}^{2}S_{\pm }+S_{\pm }S_{z}S_{\pm }S_{z}+S_{\pm }^{2}S_{z}^{2}-\kappa
S_{\pm }^{2}\right]  
\label{4-tensor} \\
T_{4,\pm 1}^{S} &=&\frac{\sqrt{7}}{2^{5/2}}\left[ S_{z}^{3}S_{\pm
}+S_{z}^{2}S_{\pm }S_{z}+S_{z}S_{\pm }S_{z}^{2}+S_{\pm }S_{z}^{3}-\kappa
\left( S_{z}S_{\pm }+S_{\pm }S_{z}\right) \right]   \\
T_{4,0}^{S} &=&\frac{\sqrt{35}}{4}\left( S_{z}^{4}-\kappa S_{z}^{2}+\lambda
\right)
\end{eqnarray}
where we have introduced notations:

\begin{equation}
\kappa =\frac{6S(S+1)-5}{7};~\lambda =\frac{3S(S+1)\left[ S(S+1)-2\right] }{%
35}  
\label{kl}
\end{equation}

\end{document}